 \documentstyle[aps,psfig]{revtex}
\begin{document}
\newcommand{\be}{\begin{equation}}
\newcommand{\ee}{\end{equation}}
\newcommand{\ba}{\begin{eqnarray}}
\newcommand{\ea}{\end{eqnarray}}
\newcommand{\nn}{\nonumber\\}
\newcommand{\half}{\textstyle{\frac{1}{2}}}
\newcommand{\quart}{\textstyle{\frac{1}{4}}}
\newcommand{\al}{{\alpha}}
\newcommand{\bpsi}{\mbox{\boldmath $\psi$}}
\newcommand{\bphi}{\mbox{\boldmath $\phi$}}
\newcommand{\bxi}{\mbox{\boldmath $\xi$}}
\newcommand{\bvarphi}{\mbox{\boldmath $\varphi$}}
\newcommand{\bnabla}{\mbox{\boldmath $\nabla$}}
\newcommand{\bjmath}{\mbox{\boldmath $\jmath$}}
\newcommand{\bPi}{\mbox{\boldmath $\Pi$}}
\newcommand{\bpi}{\mbox{\boldmath $\pi$}}
\newcommand{\br}{{\bf r}}
\newcommand{\bk}{{\bf k}}
\newcommand{\bq}{{\bf q}}

\newcommand{\JSP}[3]{{\em J.~Stat.~Phys.} {\bf {#1}}, {#2} (19{#3})}

\title{Velocity Tails for Inelastic Maxwell Models}
\author{M. H. Ernst$^1$ and R. Brito$^2$\\[2mm]
$^1$Instituut voor Theoretische Fysica\\
Universiteit Utrecht, Postbus 80006\\ 3508 TA Utrecht, The Netherlands\\[1mm]
$^2$Departamento de F\'{\i}sica Aplicada I\\ Universidad Complutense\\
28040 Madrid, Spain}

\maketitle

\vspace{1cm}
\noindent

\begin{abstract}
We study the velocity distribution function for
inelastic Maxwell models, characterized by a Boltzmann
equation with constant collision rate, independent of
the energy of the colliding particles. By means of 
a nonlinear analysis of the Boltzmann equation, 
we find that the velocity distribution function decays algebraically for 
large velocities, with exponents that are analytically calculated.
\end{abstract}

\section{Introduction}
Velocity distributions  have over-populated high energy tails in
many particle systems with in-elastic interactions, as has been
discovered theoretically, and later observed  in laboratory
experiments with granular materials \cite{physics-today}. Instead
of Gaussian tail distributions, many kinetic equations typically
predict stretched exponentials, like $\exp[-A|v|^{3/2}]$ in
driven inelastic hard sphere systems (IHS)
\cite{twan+me-gran-mat}, or even tails with a higher overpopulation, like
$\exp[ -A|v|]$ in the freely evolving IHS fluids
\cite{twan+me-gran-mat,esipov}. These theoretical predictions have been
extensively verified in Direct Monte Carlo Simulations (DSMC) of
the Boltzmann equation \cite{brey,santos}, and the stretched
exponentials have been observed in laboratory experiments with
granular matter on vibrating plates \cite{physics-today}. However,
the presence and also the absence of over-populated tails depends
very strongly on the energy input or on how the system is
thermo-statted \cite{santos,piasecki}.

Recently classes of simplified kinetic models have been studied,
so-called Maxwell models, which are characterized by a Boltzmann
equation with a collision rate that is independent of the
relative kinetic energy of the colliding particles. Maxwell
molecules are for kinetic theory, what harmonic oscillators are
for quantum mechanics, and dumb-bells for polymer physics.

Ben-Naim  and Krapivsky\cite{Ben-Naim} have presented arguments
about the non-existence of scaling solutions of the Boltzmann
equation for the one-dimensional Maxwell model, and argue in
favor of multi-time scales \cite{BN+PK-2-11}. However, Puglisi et
al. \cite{puglisi-preprint,puglisi} have found an exact scaling
solution $f(v,t) \sim (1/v^d_0(t))
\tilde{f}(v/v_0(t))$ for that equation, i.e. $\tilde{f} (c) =
(2/\pi)(1+c^2)^2$. It does have a power law tail $~ (1/c^4)$.
Puglisi et al. also solve the spatially homogeneous
Maxwell-Boltzmann equation using MC simulations in one- and two-
dimensions, and, more importantly, they show that an arbitrary
initial distribution approaches this scaling solution, with power
law tails at high energies. In two-dimensions the exponent $a$ of
the power law tails depends on the degree of in-elasticity, i.e.
on the coefficient of restitution $\alpha$. The goal of this
article is to derive these power laws from the dominant small-$k$
singularity in the Fourier transform of the velocity distribution
function.

\section{Dominant small $\lowercase{k}$ singularity}
In order to analyze the large $v$-behavior of the distribution
function, it is convenient to use the Fourier transform
$\phi(k,t)$ of $f(v,t)$. It  is the generating function of the
moments $\langle v^n\rangle _t$. If $f(v,t)$ has a tail $\sim 1/
|v|^{a+d}$, then the moments with $n>a$ are divergent, and so is
the $n$-th derivative of the generating function at $k=0$, i.e.
$\phi(k,t)$ is singular at $k=0$. Suppose the dominant small-$k$
singularity of $\phi(k,t)$ is $\sim |k|^a$, where $a$ is
different from an even integer (even powers of $k$ represent
contributions that are regular at small $k$), then the inverse
Fourier transform scales as $1/|v|^{a+d}$ at large $v$.

By applying Bobylev's Fourier transform method
\cite{bobylev-bgk,phys-rep-ME}, we obtain a nonlinear equation for
$\phi(k,t)$, and we determine its dominant small-$k$ singularity.
In doing so we have derived a transcendental equation for the
exponent $a$ in the power law tail $\sim 1/ |c|^{a+d}$  of the
scaling solutions $\tilde{f}(v/v_0)$ of the Boltzmann equation
for Maxwell molecules with in-elastic hard sphere interactions in
arbitrary dimensions.

The  Boltzmann equation for the $d$-dimensional in-elastic
Maxwell model reads,
\be \label{BEq}
\partial_t f(v_1,t) =\int_n \int dv_2 \left[ \textstyle{\frac{1}{\al}}
f(v_1^{**})f(v_2^{**}) - f(v_1)f(v_2)\right].
\ee
Here  $\int_n (\cdots) = (1/\Omega_d) \int dn(\cdots)$ is an
average over a $d$-dimensional solid angle where $\Omega_d = 2
\pi^{d/2}/ \Gamma(\half d)$. The velocities $v_i^{**}$ with
$i,j =\{1,2|\}$ denote the $d$-dimensional {\em restituting}
velocities, and $v_i^{*}$ the corresponding {\em post-collision}
velocities. They are defined as,
\ba
v_i^{**}&=& v_i -\half (1+\textstyle{\frac{1}{\al}})v_{ij}\cdot nn
\nn v_i^{*}& =& v_i -\half (1+\al)v_{ij}\cdot nn,
\ea
with $v_{ij} =v_i-v_j$, and $n$ is a unit vector along the line
of centers of the interacting particles. In one-dimension, the
tensorial product $nn$ can be replaced by $1$. From the
normalization of $f$ it follows that the loss term reduces to $-
f(v_1,t)$, i.e. the collision frequency is unity, and the
dimension-less time $t$ counts the average number of collisions
per particle.

We first illustrate the method for the one-dimensional case.
Fourier transformation of the Boltzmann equation yield then,
\be
\partial_t \phi(k,t)= \phi(p k,t)\phi((1-p)k,t)-\phi(k,t),
\ee
where we have used that $\phi(0,t)=1$ and $p=\half (1+\al)$. The
equation for the scaling solution, $\phi(k,t)= \Phi(v_0(t)k)$
simplifies to,
\be
-\gamma k d\Phi(k)/dk +\Phi(k)=\Phi(p k)\Phi((1-p)k),
\label{fourier-BE}
\ee
where the exponent $\gamma$ in $v_0(t) =v_0(0)\exp[-\gamma t]$ is
still to be determined.

The requirement that the total energy be finite, imposes the the
lower bound $a>2$ on the exponent. We therefore make the ansatz
that the dominant small-$k$ singularity has the form,
\be
\Phi(k) = 1- \half \langle(k \cdot c)^2\rangle  + A|k|^a.
\ee
Inserting this in (\ref{fourier-BE}), and equating the
coefficient of equal powers of $k$ yields the equation,
\be
a = \frac{1-p^a-(1-p)^a}{p(1-p)}.
\ee
The smallest root of this equation, satisfying $a>2$, is $a=3$,
and $A$ is left undetermined.  Consequently the scaling solution
has a power law tail, $\tilde{f}(c) \sim 1/c^4$.

The same method can be applied to the $d$-dimensional case.
Application of  Bobylev's Fourier transform method to the
Boltzmann equation for this case\cite{bobylev,bobylev-bgk} yields
the transformed equation,
\be
\partial_t \phi(k,t)= \int_n \phi(k_+,t)\phi(k_{-},t)-\phi(k,t),
\ee
where the $n$-average is defined below (1), and
\ba
k_{+} &= p k \cdot n n  \qquad & |k_+|^2 = p^2 k^2 (\hat{k}\cdot
n)^2
\nn k_{-} & =k - k_{+} \qquad & |k_{-}|^2 = k^2 [ 1- q (\hat{k}
\cdot n)^2],
\ea
where $q=p(2-p)$ is a positive number. We proceed in the same way
as in the one-dimensional case, and obtain the equation for the
scaling solution,
\be
-\gamma k d\Phi(k)/dk +\Phi(k)= \int_n \Phi( k_+)\Phi(k_{-}).
\ee

\begin{figure}[h]
$$\psfig{file=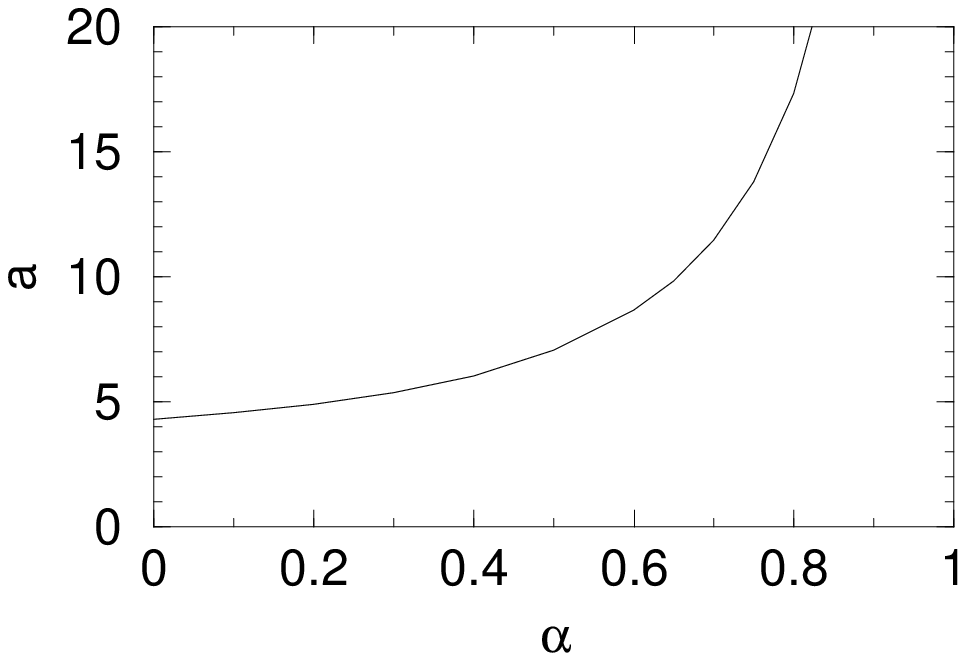,height=5cm}\qquad \psfig{file=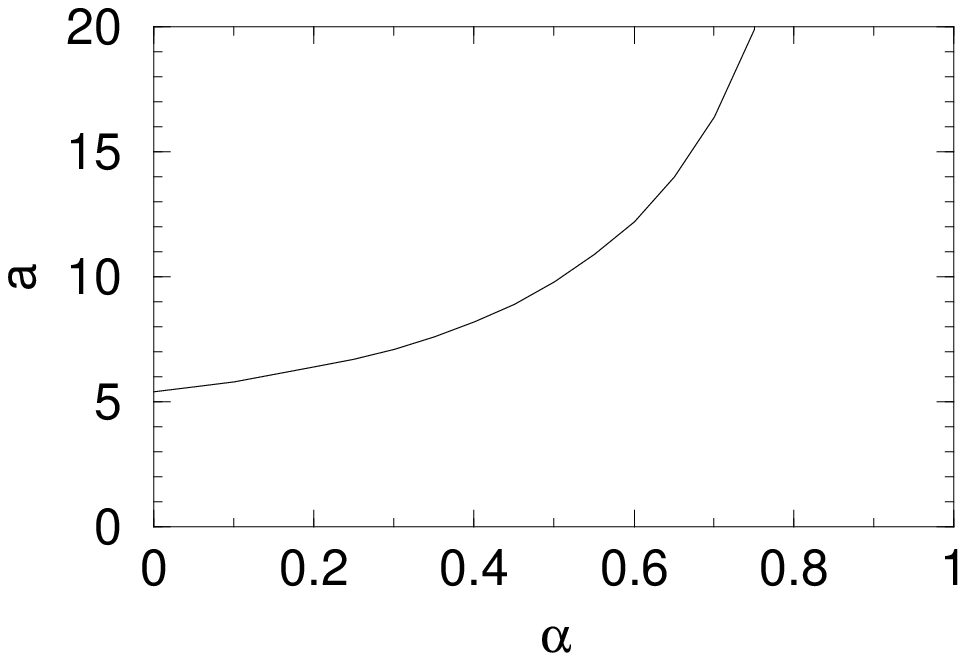,height=5cm}$$
\caption{Solution of Eq.(\protect{\ref{self-consist-a}})
as a function of $\alpha$ for 2 diemnsions (left panel)
and 3 dimensions (right panel). The solution  diverges as $\alpha\to 1$ (elastic
limit), because $\tilde f$ becomes Maxwellian in this limit. We note
that $a=2$ always satisfies Eq.(\protect{\ref{self-consist-a}}), and
the solution shown here is the one different from $a=2$.}
\label{figure2d}
\end{figure}

\noindent
Inserting the ansatz (5) into (8), and equating the coefficients
of equal powers of $k^s$  yields,
\be
a \gamma \langle |k \cdot c|^a \rangle = \int_n  
\langle |k \cdot c|^a - |k_+ \cdot
c|^a - |k_{-} \cdot c|^a \rangle,
\ee
for $s=2,a$. In order to carry out the angular $\hat{c}$-average
in $\langle |q \cdot c|^a\rangle $ 
with $q= \{k,k_+,k_{-}\}$ we choose $q$ as polar axis, and
denote $q \cdot c = q\;c\; \hat{q} \cdot \hat{c} = q\;c\; \cos
\theta$, then $\langle |q \cdot c|^a \rangle = |q|^a \langle |c|^a \rangle K^{(d)}_a$, where
$K^{(d)}_a$ is the average of $|\cos \theta|^a$ over a
$d$-dimensional solid angle, which equals
\be
K^{(d)}_a  =
\Gamma(\half(a+1))\;\Gamma(\half d)/
\Gamma(\half (a+d))\;\Gamma(\half),
\ee
where  $ q =p(2-p)$. Finally we carry out the angular
$n$-averages using (8), and obtain,
\ba
\int_n |k_+|^a &=& k^ap^a K^{(d)}_a  \nn
\int_n |k_{-}|^a &=& k^a \int_n [1-q(\hat{k}
 \cdot n)^2]^{a/2}= k^a L^{(d)}_a (q).
\ea
Insertion of these results in (10) for a=2, yields
\be
\gamma = \textstyle{\frac{1}{d}} p(1-p) = \textstyle{\frac{1}{4d}} (1-\al^2).
\ee
For the exponent $a$, featuring in the power law tail of the
scaling function $\tilde{f} (c) \sim 1/c^{a+d}$, we obtain the
transcendental equation,
\be\label{self-consist-a}
a= \frac{1-p^a K^{(d)}_a -L^{(d)}_a(q)}{\textstyle{\frac{1}{d}} p(1-p)}.
\ee
The two most interesting cases are $d=2,3$, where
\ba
L^{(2)}_a (q) &=& \frac{2}{\pi} \int^{\pi/2}_0 d \theta
[1-q\cos^2 \theta]^{a/2}
\nn L^{(3)}_a (q)&=& \int^1_0 d x [1-q x^2]^{a/2},
\ea
and one can verify that $a=2$ is also a solution of (14). We look
for the smallest solution $a(\al)$ of this transcendental
equation with $a>2$. The numerical solutions for $d=2,3$ are shown
in Figure \ref{figure2d} as a function of $\alpha$. If $p=\half (1+\al)
\uparrow 1$ the root $a(\al)$ moves to $\infty$, as it should, which
is consistent with a Maxwellian tail distribution for the elastic
case.

The simulations in Ref.\cite{puglisi-preprint} of the
two-dimensional Maxwell - Boltzmann equation show for the exponent
$a({\al=0})+2 \simeq 5$,where our analytical method predicts
$a({\al=0})+2 \simeq 6.2$.  In fact, closer inspection of their
two-dimensional scaling plot shows that the slope of their
log-log plot  of $\tilde{f} (c)$ versus $c$ increases at larger
velocities, approaching the exact prediction of the Boltzmann
equation. However, at these large velocities the statistical
errors in their simulations are too large to make a quantitative
comparison for larger $\al$-values.

\end{document}